\documentclass[aps,twocolumn,groupedaddress,amsmath,amssymb]{revtex4}
\usepackage{amsmath, dsfont}
\usepackage{amssymb}
\usepackage[dvips]{graphics}
\usepackage{epsfig}
\usepackage{calc}
\usepackage{amsfonts}
\usepackage{graphicx}
\usepackage[ansinew]{inputenc}

\begin{document}
\setcounter{page}{1}
\title{Thermodynamics of charged Lifshitz black holes with quadratic corrections}

\author{Mois\'es Bravo-Gaete}
\email{mbravog-at-inst-mat.utalca.cl} \affiliation{Instituto de
Matem\'atica y F\'isica, Universidad de Talca, Casilla 747, Talca,
Chile}

\author{Mokhtar Hassa\"ine}
\email{hassaine-at-inst-mat.utalca.cl} \affiliation{Instituto de
Matem\'atica y F\'isica, Universidad de Talca, Casilla 747, Talca,
Chile}

\begin{abstract}
In arbitrary dimension, we consider the Einstein-Maxwell Lagrangian supplemented by the more general quadratic-curvature corrections. For this model, we derive four classes of charged Lifshitz black hole solutions for which the metric function is shown to depend on a unique integration constant. The masses of these solutions are computed using the quasilocal formalism based on the relation established between the off-shell ADT and Noether potentials. Among these four solutions, three of them are interpreted as extremal in the sense that their mass vanishes identically. For the last family of solutions, the quasilocal mass and the electric charge both are shown to depend on the integration constant. Finally, we verify that the first law of thermodynamics holds for each solution and a Smarr formula is also established for the four solutions.

\end{abstract}

\maketitle

\section{Introduction}
During the last decade, there has been a certain interest in extending the
ideas underlying the standard relativistic AdS/CFT
correspondence \cite{Maldacena:1997re} to non-relativistic physics in order to gain a better understanding
of the condensed matter physics, and particulary to physical systems that exhibit a
dynamical scaling near fixed points. These latter are characterized by an invariance
under a rescaling symmetry with different weights between
the space and the time that reads
\begin{eqnarray}
t\to\lambda^z\,t,\qquad\qquad\qquad\qquad \vec{x}\to\,\lambda \vec{x}.
\label{anisyymetry}
\end{eqnarray}
The constant $z$ which is called the dynamical exponent reflects the
anisotropic symmetry. In analogy with the AdS case, the gravity dual metric in $D-$dimensions refereed as the Lifshitz metric \cite{Kachru} is given
by
\begin{equation}
\label{Lifmetric} ds_{{\cal
L}}^2=-\left(\frac{r}{l}\right)^{2z}dt^{2}+\frac{l^{2}}{r^{2}}\,dr^{2}+\frac{r^{2}}{l^{2}}\,\sum_{i=1}^{D-2}dx_{i}^{2},
\end{equation}
and, it is easy to see that the anisotropic
scaling transformation (\ref{anisyymetry}) together with the
rule $r\to \lambda^{-1}r$ act as an isometry for this metric.

Soon after the introduction of the Lifshitz background (\ref{Lifmetric}), it was realized that, in contrast
with the standard AdS case $z=1$, these metrics are not solutions of the vacuum Einstein equations, and instead require the
introduction of some matter source or to consider higher-order curvature terms \cite{Kachru}. Finite temperature effects are introduced via
Lifshitz black holes which refer to black hole metrics whose
asymptotic behavior reproduces the Lifshitz background (\ref{Lifmetric}). Up to now, a relatively important variety of Lifshitz black hole solutions have been reported in the current literature. For examples, for the three-dimensional new massive gravity theory introduced in \cite{Bergshoeff:2009hq}, there exists a Lifshitz black hole solution with a dynamical exponent $z=3$, \cite{AyonBeato:2009nh}. Higher-dimensional generalizations of this vacuum solution have been derived in \cite{AyonBeato:2010tm} for more general quadratic corrections of the Einstein gravity. In four dimensional conformal gravity, Lifshitz black holes were constructed for dynamical exponents $z=0$ and $z=4$ in \cite{Lu:2012xu} and their Maxwell electrically charged versions were recently reported in \cite{Fan:2014ala}. Lifshitz black holes for cubic theories were investigated in \cite{Giacomini:2012hg} as well as for contractions of the Weyl tensor \cite{Oliva:2012zs}. Examples of charged Lifshitz black hole solutions have been obtained in presence of Abelian sources \cite{Taylor,DanielssonI}, in the case of a Maxwell-Proca theory \cite{Pang:2009pd} or more generally for nonlinear electrodynamics source \cite{Alvarez:2014pra}. Scalar fields can also accommodate Lifshitz black hole metrics in the case of the type IIB Sugra with a dynamical exponent $z=3/2$, \cite{Azeyanagi:2009pr} or with a nonminimal coupling parameter, see Refs. \cite{Correa:2014ika} and \cite{Ayon-Beato:2015jga} or in the case of a particular truncation of the Horndeski theory \cite{Bravo-Gaete:2013dca}.

Although, Lifshitz black holes can be found analytically, and their temperature as well as their entropy can be easily computed, their thermodynamics issue remains an hard task. Indeed, the definition of Lifshitz mass, because of the rather unconventional falloff behavior, is far less clear than that in the AdS case. Also, an important class of Lifshitz solutions have been obtained in the case of higher-derivative terms which render difficult their linearization analysis. Recently, it has been proposed a novel way of obtaining quasilocal charges for black hole solutions of any gravity theory invariant under diffeomorphism, see \cite{Kim:2013zha} and \cite{Gim:2014nba}. The main result of these two last references is the fact that the off-shell ADT potential \cite{Deser:2002rt, Deser:2002jk, Senturk:2012yi} can be expressed in terms of the off-shell Noether potential which is only build in the basis of the Lagrangian and the Killing vectors associated to the conserved charges without the need of considering the linearization of the field equations. Another important feature of this method lies in the fact that it may apply for arbitrary asymptotic spacetime, and not necessarily for asymptotically flat of AdS spacetimes. For example, this formalism has been shown to be efficient in order to obtain the correct mass of the $z=3$ Lifshitz black hole solution of new massive gravity \cite{AyonBeato:2009nh}, see \cite{Gim:2014nba}. More recently, the authors in Ref. \cite{Ayon-Beato:2015jga}  have considered the model of new massive gravity in three dimensions with a nonminimal scalar field for which they obtained three classes of Lifshitz black hole solutions. The masses of these solutions computed within the quasilocal formalism fit perfectly with the first law of thermodynamics, and as an additional check they showed that these expressions of the masses are in accordance with the anisotropic Cardy formula \cite{Gonzalez:2011nz} assuming that the ground state is played by the respective soliton \cite{Ayon-Beato:2015jga} .

In the present paper, we further explore this quasilocal formalism in the case of charged Lifshitz black holes. In arbitrary dimensions $D$, we consider the Einstein-Maxwell Lagrangian supplemented by a cosmological constant with the quadratic corrections build from the scalar curvature $R$, the Ricci tensor $R_{\mu\nu}$ and the Riemann tensor $R_{\mu\nu\rho\sigma}$ as
\begin{eqnarray}
S=&&\frac{1}{2 \kappa}\,\int{d}^Dx\sqrt{-g}
\left(R-2\Lambda+\beta_1{R}^2
+\beta_2{R}_{\alpha\beta}{R}^{\alpha\beta} \right.\nonumber\\
&&+\left.\beta_3{R}_{\alpha\beta\mu\nu}{R}^{\alpha\beta\mu\nu}
\right)-\frac{1}{4}\int{d}^Dx\sqrt{-g}\,F_{\alpha\beta}F^{\alpha\beta},\nonumber\\
=&&\int{d}^Dx\sqrt{-g}\,{\cal L},
\label{eq:Squad}
\end{eqnarray}
where $F_{\alpha\beta}=\partial_{\alpha}A_{\beta}-\partial_{\beta}A_{\alpha}$
corresponds to the field strength of the Maxwell field, and where the $\beta_i$ are constants. The field equations obtained varying
the action (\ref{eq:Squad}) read
\begin{subequations}
\label{eqs}
\begin{eqnarray}
&&{E_{\mu \nu}:=G_{\mu\nu}+\Lambda g_{\mu\nu}+K_{\mu\nu}-
\kappa \,T_{\mu \nu}=0},\label{eq:motionsmetric}\\
&&{\nabla_{\mu} F^{\mu\nu}=0} \label{eq:maxwell},
\end{eqnarray}
where
\begin{eqnarray}
T_{\mu \nu}&=& F_{\mu\sigma}F^{\sigma}_{\nu}-
\frac{1}{4}g_{\mu\nu}F_{\alpha\beta}F^{\alpha\beta},
\\
K_{\mu\nu}&=&\left(\beta_2+4\beta_3\right)\square{R}_{\mu\nu}+\frac12\left(4\beta_1+\beta_2\right)g_{\mu\nu}\square{R}
\nonumber\\
&-&\left(2\beta_1+\beta_2+2\beta_3\right)\nabla_\mu\nabla_\nu{R}+2\beta_3R_{\mu\gamma\alpha\beta}R_{\nu}^{~\gamma\alpha\beta}
\nonumber\\
&+&2\left(\beta_2+2\beta_3\right)R_{\mu\alpha\nu\beta}R^{\alpha\beta}
-4\beta_3R_{\mu\alpha}R_{\nu}^{~\alpha}+2\beta_1RR_{\mu\nu}
\nonumber\\
&-&\frac12\left(\beta_1{R}^2+\beta_2{R}_{\alpha\beta}{R}^{\alpha\beta}
+\beta_3{R}_{\alpha\beta\gamma\delta}{R}^{\alpha\beta\gamma\delta}
\right)g_{\mu\nu}.
\end{eqnarray}
\end{subequations}
We are interested on looking for charged black hole solutions of the
field equations (\ref{eqs}) with a purely electric ansatz such that asymptotically the metric
behaves as the Lifshitz spacetime (\ref{Lifmetric}). In doing so, we
opt for the following Lifshitz type ansatz
\begin{eqnarray}\label{metric}
ds^{2}_{D}=-\left(\frac{r}{l}\right)^{2z}f(r)\,dt^{2}+\frac{l^{2}}{r^{2}}\,\frac{dr^{2}}{f(r)}+\frac{r^{2}}{l^{2}}\,\sum_{i=1}^{D-2}
dx_{i}^{2},
\end{eqnarray}
where we require the unknown metric function $f(r)$ to satisfy
$\lim_{r\to +\infty}f(r)=1$, and the Maxwell potential to be of the form $A_{\mu}dx^{\mu}=A_t(r)dt$.

The plan of the paper is organized as follows. In the next section, we show that the metric function within our ansatz (\ref{metric}) behaves generically as
$f(r)=1-M\left(\frac{l}{r}\right)^{\chi}$,
where $M$ is an integration constant, and where $\chi$ is a positive constant that always depends on the dimension $D$. Using this result together
with the integration of the Maxwell equation (\ref{eq:maxwell}), we write down the generic formula of the quasilocal mass associated to the Killing vector $\xi^{\mu}\partial_{\mu}=\partial_t$. In Sec. III, we exhibit explicitly the four families of charged Lifshitz black hole solutions, and for each of them we compute the mass, the electric charge, the entropy and verify that the first law of thermodynamics holds in each case. In Sec. IV, we discuss the lower-dimensional cases, $D=3$ and $D=4$ while in the last section, we report our conclusions. Finally, an appendix is added for clarity where we give some of the formulas needed to derive our results.

\section{Quasilocal mass, entropy and electric charge}
In this section, we will show that the metric function within our ansatz (\ref{metric}) behaves generically as $f(r)=1-M\left(\frac{l}{r}\right)^{\chi}$
and with this result together with the integration of the Maxwell equation (\ref{eq:maxwell}), we will express in a generic way
the expression of the quasilocal mass. The details concerning the  explicit solutions will be given in the next section.

Considering the metric ansatz (\ref{metric}), the following
combination of the Einstein equations (\ref{eq:motionsmetric})
\begin{eqnarray}
-\frac{l^{4}}{f(r)}\,\left( E_{t}^{t}-E_{r}^{r}\right)=0,
\label{comb1}
\end{eqnarray}
yields a fourth-order Cauchy differential equation for the unknown metric function $f(r)$,
which is given in the appendix (\ref{eq:cauchy}), and whose solution is generically expressed as
\begin{eqnarray}
f(r)=1-\sum_{i=1}^{4} M_{i}\left(\frac{l}{r}\right)^{\alpha_{i}},
\label{f1}
\end{eqnarray}
where the $M_{i}$ are some integration constants and the $\alpha_{i}$ are the roots of
the characteristic polynomial. Before following this generic analysis, we would like to stress two things. Firstly, there also exists the
possibility of having multiple roots for which the metric function $f(r)$ will involve logarithmic contributions, but in those cases, the
remaining independent Einstein equations will impose that all the integration constants associated to the logarithmic contributions must be zero. Secondly,
in the special point where the quadratic corrections yield to the Gauss-Bonnet density, namely $\beta_{1}=-\frac{1}{4}\beta_{2}=\beta_{3}$, the fourth-order Euler equation reduces to the following constraint
\begin{eqnarray*}
{\frac {f(r)  \left[ {\beta_2}\, \left( D-3 \right) \left( D-4
\right) f(r)  +2\,{l}^{2} \right] \left( z-1 \right)  \left( D-2
\right) }{{l}^{4}}}=0,
\end{eqnarray*}
which in turns implies that the dynamical exponent must be restricted to $z=1$. In this particular case, the charged black hole solutions are asymptotically AdS, and these solutions have already been studied in
Refs. \cite{Banados:1993ur,Crisostomo:2000bb,Aros:2000ij}. However, in our case, we are only interested on the purely Lifshitz black holes, that means $z\not=1$. Following our analysis, after a tedious computation, one can show that the combination of the Einstein equations  $E_{t}^{t}+E_{i}^{i}=0$ imposes that only one of the integration constants of the metric function (\ref{f1}) is not zero and fixes conveniently the constants $\beta_1$, $\beta_3$ and $\Lambda$. This means that the metric solution can be generically written as
\begin{equation}\label{metricfunction}
f(r)=1-M\,\left(\frac{l}{r}\right)^{\chi},
\end{equation}
where $M$ is an integration constant, and the combination $E_{t}^{t}+E_{i}^{i}=0$ becomes simply given by
\begin{eqnarray}
\left( 2-2\,z+\chi \right) {M}^{2} \left( \chi+2-D \right) \left(
-z +2-D+2\,\chi \right)&&\nonumber\\
\times \Xi_{1}\,\left(\frac{r}{l}\right)^{-2\,\chi} +
\left(-2\,D+4+\chi\right)\,M\,\Xi_{2}\,\left(\frac{r}{l}\right)^{-\chi}=0,\label{eq:ettii}&&
\end{eqnarray}
where $\Xi_{1}$ and $\Xi_{2}$ are constants whose expressions are given in the appendix. Finally, the remaining independent Einstein equation $E_{t}^{t}=0$ can be casted in the following form
\begin{eqnarray}
E_{t}^{t}&=&-\frac{1}{2}\, \left( D-2 \right) \left( 2-2\,z+\chi
\right) \left( 2\,\chi-D+2-z \right) \nonumber\\
&\times& \Xi_{1}{M}^{2} \left( {\frac {r}{l}} \right) ^{-2\, \chi}
-\left( D-2 \right) \Xi_{2} \,M \left( {\frac {r}{l}} \right)
^{-\chi}\nonumber\\
&+&\frac{1}{2}\,\kappa\, {Q}^{2}\,\left( {\frac {r}{l}} \right)
^{-2(D-2)}=0,
\label{Ett=0}
\end{eqnarray}
where we have explicitly used the expression of the purely electrical Maxwell field solution of the equation (\ref{eq:maxwell})
\begin{equation}
F_{rt}=Q\,\left(\frac{r}{l}\right)^{z-D+1},
\label{maxsol}
\end{equation}
where $Q$ is an integration constant. From equations
(\ref{eq:ettii}-\ref{Ett=0}), it is now simple to see that there
exist four classes of solutions for the ansatz metric (\ref{metric})
with a metric function given by (\ref{metricfunction}), and these
are summarized as follows
\begin{eqnarray}
\begin{tabular}{|l|l|l|l|l|}
\hline Family & $Q\qquad $ & $\chi\qquad $ & $z\qquad $ & Extra \qquad\\
\hline A & $\propto M$ & $D-2$ & Free & $\Xi_{2}=0$ \\
\hline B & $\propto \sqrt{M}$ & $2(D-2)$ & Free & $\Xi_{1}=0$ \\
\hline C & $\propto \sqrt{M}$ & $2(D-2)$ & $D-1$ &  \\
\hline D& $\propto \sqrt{M}$ & $2(D-2)$ & $3(D-2)$ &  \\
\hline
\end{tabular}
\label{sol4}
\end{eqnarray}
The details concerning these four solutions will be reported in the next section.

We are now in position to write down the expression of the quasilocal mass for the generic solutions described by the ansatz metric (\ref{metric}) with a metric function given by (\ref{metricfunction}) together with the form of the Maxwell field (\ref{maxsol}). And interestingly enough, we will show that these four classes of solutions may also emerge from the expression of the quasilocal mass.

As said in the introduction, the main result of the authors in Refs.
\cite{Kim:2013zha,Gim:2014nba} lies in the following relation
$$
\sqrt{-g}\,Q_{\mbox{\tiny{ADT}}}^{\mu\nu}=\frac{1}{2}\delta
K^{\mu\nu}-\xi^{[\mu}\Theta^{\nu]},
$$
that allows to express the off-shell ADT potential
$Q_{\mbox{\tiny{ADT}}}^{\mu\nu}$ \cite{Deser:2002rt,Deser:2002jk} to
the off-shell Noether potential $K^{\mu\nu}$. In this relation,
$\xi^{\mu}\partial_{\mu}$ denotes the Killing vector field which in
our case is $\partial_t$, and $\Theta^{\mu}$ represents a surface
term arising from the variation of the action. For the model
considered here, these expressions are given by
\begin{eqnarray}
\Theta^\mu (\delta g,\delta A)&=&2\sqrt{-g}\Big[P^{\mu(\alpha
\beta)\gamma}\nabla_\gamma\delta g_{\alpha\beta} -\delta
g_{\alpha\beta}\nabla_\gamma
P^{\mu(\alpha\beta)\gamma} \nonumber\\
&-&\frac{1}{2}\,F^{\mu \nu}\,\delta A_{\nu}\Big], \label{eq:theta}\\
K^{\mu\nu}
&=&\sqrt{-g}\,\Big[2P^{\mu\nu\rho\sigma}\nabla_\rho\xi_\sigma
-4\xi_\sigma\nabla_\rho
P^{\mu\nu\rho\sigma}\nonumber\\
&+&F^{\mu \nu} \xi^{\sigma} A_{\sigma}\Big] \label{eq:K},
\end{eqnarray}
with $P^{\mu \nu \sigma \rho}=\frac{\partial \mathcal{L}}{\partial
R_{\mu \nu \sigma \rho}}$, where $\mathcal{L}$ is the Lagrangian
defined in (\ref{eq:Squad}). As shown before, the solution depends
continuously on a constant $M$, and hence in order to define the
conserved charge in the interior region and not in the asymptotic
region a parameter $s$ with range $s\in [0,1]$ is introduced as
$sM$. The advantage of this re-definition lies in the fact that it
allows to interpolate between the free parameter solution $s=0$ and
the solution with $s=1$. In doing so, the quasi-local charge is
defined as
\begin{equation}
\label{eq:charge} {\cal M}(\xi)\!=\int_{\cal B}\!
d^{D-2}x_{\mu\nu}\Big(\Delta K^{\mu\nu}(\xi)-2\xi^{[\mu} \!\!
\int^1_0ds~ \Theta^{\nu]}(\xi\, |\, s)\Big),
\end{equation}
where $\Delta K^{\mu\nu}(\xi) \equiv
K^{\mu\nu}_{s=1}(\xi)-K^{\mu\nu}_{s=0}(\xi)$, denotes the variation
of the Noether potential from the vacuum solution, and
$d^{D-2}x_{\mu\nu}$ represents the integration over the compact co-dimension
two-subspace. In the present case, this last expression becomes
\begin{widetext}
\begin{eqnarray}
{\cal M}(\xi)\!=\Big[\frac{M^{2}\,\Psi_{1}}{2
\,\kappa}\,\frac{r^{D-2+z-2
\chi}}{l^{D+1+z-2\chi}}+\frac{M\,\Psi_{2}
}{2\,\kappa}\,\frac{r^{D-2+z-\chi}}{l^{D+1+z-\chi}}-\frac{Q^{2}}{2\left(z-D+2\right)}\,\frac{r^{z-D+2}}{l^{z-D+1}}\Big]\Omega_{D-2},\label{eq:mastereq}
\end{eqnarray}
\end{widetext}
where $\Psi_{1}$ and $\Psi_{2}$ are constants reported in the
appendix and  $\Omega_{D-2}$ represents the finite contribution of
the $(D-2)$-dimensional integration over the planar variables.
Note that the value $z=D-2$ must be excluded from this analysis;
however, in this case, one can show that the solution becomes
uncharged $Q=0$, and reduces to the vacuum solution reported in
\cite{AyonBeato:2010tm}. It is clear that the expression in the
right-hand side must not depend on the radial coordinate $r$ and
this corresponds precisely to the four classes of solutions
(\ref{sol4}) as we will show explicitly in the next section.

Since we will be interested on the thermodynamics properties of the solutions, the entropy of the solutions will be computed through the Wald formula
\begin{equation}
 S_{W} = - 2 \pi \, \Omega_{D-2} \, \left(\frac{r_{h}}{l}\right)^{D-2} \,
 \Big[ \frac{\delta {\mathcal L}}{\delta R_{abcd}}
 \, \varepsilon_{ab} \,  \varepsilon_{cd} \Big]_{r = r_{h}}, \label{w2}
\end{equation}
where $r_h$ denotes the location of the horizon. On the other
hand, the Hawking temperature reads
\begin{equation}
 T=\frac{r_{h}^{z+1}} {4 \pi\,l^{z+1} }
 f^{\prime}(r_{h}). \label{T}
\end{equation}
In order to see that our expressions of the masses fit with the first law of the thermodynamics, we also need the expression of the electric charge $\mathcal{Q}$ that is generically given by
\begin{eqnarray}\label{eq:electriccharge}
\mathcal{Q}=\int d {\Omega_{D-2}} \left(\frac{r}{l}\right)^{D-1-z}
F_{rt}.
\end{eqnarray}

\section{Four classes of charged Lifshitz black hole solutions}
We now report in details the four classes of charged Lifshitz black
hole solutions (\ref{sol4}). In each case, we compute the mass
${\cal M}$ through the formula (\ref{eq:charge}), the entropy
$S_{W}$ (\ref{w2}), the temperature $T$, the electric charge
$\mathcal{Q}$ as well as the electric potential $\Phi=-A(r_h)$.
Having these quantities in hand, we verify that the first law of
thermodynamics
\begin{eqnarray}
d{\cal M}=TdS_{W}+\Phi d\mathcal{Q}, \label{ftlwa}
\end{eqnarray}
holds for each of the solutions. We show that among our four solutions, three of them have the peculiarity that the mass vanishes identically; these latter can be interpreted as extremal charged Lifshitz black holes as those recently given in Ref. \cite{Liu:2014dva}.

\subsection{Extremal charged solution with arbitrary dynamical exponent}
The first solution holds for an arbitrary value of the dynamical exponent, and is given by
\begin{eqnarray}
ds^{2}_{D}&=&-\left(\frac{r}{l}\right)^{2z}f(r)\,dt^{2}+\frac{l^{2}}{r^{2}}\,\frac{dr^{2}}{f(r)}+\frac{r^{2}}{l^{2}}\,\sum_{i=1}^{D-2}
dx_{i}^{2}, \nonumber\\
f(r)&=&1-{M}\left(\frac{l}{r}\right)^{D-2},\label{eq:metricsoln1} \nonumber \\
F_{rt}(r)&=& {M} \Sigma_{1}
\,\left(\frac{r}{l}\right)^{z-D+1}\label{eq:maxwellsoln1},
\end{eqnarray}
where
\begin{eqnarray*}
\big(\Sigma_{1}\big)^{2}&=&\big\{\left( 2\,z-D \right)  \big[
2\,{z}^{2}- \left( z-1 \right) \left( D-2 \right)  \big]
 \nonumber\\
&\times& \left( z-D+2 \right) \left( z+D-2 \right) \left( D-2
\right) \big\}\nonumber\\
&\Big{/}&{[4\,\kappa\, l^{2} z P_{3}(z)]},
\end{eqnarray*}
and
\begin{equation*}
P_{3}(z)=2\,{z}^{3}-2\,{z}^{2} \left( 2\,D-3 \right) - \left( D-2
\right) \big[ z \left( D-9 \right) +4 \big] .
\end{equation*}
In this case, the coupling constants of the theory are tied as follows
\begin{eqnarray*}
\beta_{1}&=&l^{2}\,\big[2 \left( D-2 \right) {z}^{5}- \left(
3\,{D}^{2}-4\,D-8 \right) {z}^ {4} \nonumber\\
&+&\left( 10\,{D}^{2}-36\,D+28 \right) {z}^{3}- \left( D-2 \right)
\nonumber\\
&\times&\left( {D}^{2}+11\,D-32 \right) {z}^{2}+ \left( 2\,D-5
\right) \left( D-2 \right)
\nonumber\\
&\times&\left( D+2 \right) z-2\, \left( D-2 \right) ^{3}\big]\nonumber\\
&\Big{/}&\big[2\,{z}^{2} \left( D-2 \right)\left( D-3 \right) \left(
D-4 \right)P_{3}(z)\big], \\ \nonumber\\
\beta_{2}&=&l^{2}\,\big\{\left[ 2\,{z}^{2}-z \left( D-2 \right) +D-2
\right]
\left[ -2\left( D-2 \right) {z}^{3}\right.\nonumber\\
&+&\left. D z \left( z-1 \right) \left( 3\,D-7 \right) + \left(
D-2\right) \left( {D}^{2}-3\,D+4 \right)
\right]\big\}\nonumber\\
&\Big{/}& \big[\left( D-2 \right)  \left( D-4 \right)    \left( D-3
\right) {z}^{2} P_{3}(z)\big],\\ \nonumber \\
\beta_{3}&=&\frac{l^{2}\big[2\,{z}^{2}- \left( z-1 \right)  \left(
D-2 \right)\big]}{4\, \left( D-3 \right)  \left( D-4 \right)
{z}^{2}}, \\ \nonumber\\
\Lambda&=&-\frac{\left( D-2 \right)  \big[ {z}^{2}+ \left( 2\,z-1
\right)  \left( D-2 \right)  \big] }{4\,z\,l^{2}}.
\end{eqnarray*}
Plugging this solution in the expression (\ref{eq:mastereq}) implies that the mass ${\cal M}$ vanishes identically. On the other hand, the Wald entropy is no zero and given by
\begin{eqnarray}
S_{W}&=&{\frac {4\,\pi\,\Omega_{D-2}\,
\big(\Sigma_{1}\big)^{2}\,r_{h}^{D-2}}{ \left( z -D+2 \right) \left(
D-2 \right)\,l^{D-4} }} ,
\end{eqnarray}
while the Hawking temperature is
\begin{eqnarray}
T={\frac { \left( D-2 \right) r_{h}^{z}}{4\,\pi\,l^{z+1} }}.
\end{eqnarray}
The charge and the electric potential read respectively
\begin{eqnarray}
\mathcal{Q}= \Sigma_{1}\,\Omega_{D-2}
\left(\frac{r_{h}}{l}\right)^{D-2},
\end{eqnarray}
and
\begin{eqnarray}
\Phi=-A(r_{h})=-{\frac
{\Sigma_{1}\,r_{h}^{z}}{\big(z-D+2\big)\,l^{z-1}}}.
\end{eqnarray}
It is a matter of check to see that in spite of having a zero mass, the first law (\ref{ftlwa}) still holds in the form
\begin{equation*}
d \mathcal{M}=0=T d S_{W}+\Phi d \mathcal{Q},
\end{equation*}
and hence, this solution can be interpreted as an extremal charged Lifshitz black hole.

\subsection{Second extremal family with an arbitrary $z$}
A second extremal family of solution is found when $\chi=2(D-2)$ and for an arbitrary value of the dynamical exponent,
\begin{eqnarray}
ds^{2}_{D}&=&-\left(\frac{r}{l}\right)^{2z}f(r)\,dt^{2}+\frac{l^{2}}{r^{2}}\,\frac{dr^{2}}{f(r)}+\frac{r^{2}}{l^{2}}\,\sum_{i=1}^{D-2}
dx_{i}^{2}, \nonumber\\
f(r)&=&1-M\left(\frac{l}{r}\right)^{2(D-2)},\label{eq:metricsoln2} \nonumber\\
F_{rt}&=& \sqrt{M}\,\Sigma_{2}\,\left(\frac{r}{l}\right)^{z-D+1},
\label{eq:maxwellsoln2}
\end{eqnarray}
where
\begin{eqnarray*}
\big(\Sigma_{2}\big)^{2}&=&\big\{\big[ 2\,{z}^ {2}+ \left( D-2
\right) \left(
D-1-4\,z \right) \big] \nonumber\\
&\times& \left( D-2 \right)  \left( z+2\,D-4 \right) \left( z-D +2
\right)\, \big\}\nonumber\\
&\Big{/}&{[2\,\kappa\,l^{2} z \, P_{2}(z)]},
\end{eqnarray*}
with
\begin{equation*}
P_{2}(z)= {z}^{2}- \left( D-2 \right) \left( D-4+ 2\,z \right).
\end{equation*}
The space of parameters is defined by
\begin{eqnarray*}
\beta_{1}&=& l^{2} \big[2 \left( D-2 \right) {z}^{4}+ \left(
25\,D-7\,{D}^{2}-20 \right) {z }^{3} \nonumber\\
&+&2\, \left( D-2 \right) \left( 3\,{D}^{2}-8\,D+3 \right) {z}^{2 }-
\left( D-2 \right)
\nonumber\\
&\times& \left( 3\,{D}^{2}-5\,D-4 \right) z -2\, \left( D-4 \right)
\left( D-2 \right)
^{3}\big]\\
&\Big{/}&\big[4 \left( D-2 \right)  z \left( D-3 \right) \left( D-4
\right) \left( z-D+2 \right) P_{2}(z) \big],\nonumber\\
\nonumber\\
\beta_{2}&=&l^{2} \big\{\left[  -4\left( D-2 \right) {z}^{2}+ \left(
20-25\,D+7\,{D}^{2} \right) z \right.\nonumber\\
&+&\left.2\, \left( D-1 \right) \left( D-2\right)  \left( D-4
\right) \right] \nonumber\\
&\times& \left[ 2\,{z}^{2}+ \left(
D-2 \right)  \left( D-1- 4\,z \right) \right]\big\}\\
&\Big{/}& \big[4\, \left( D-4 \right) z \left( z-D+2 \right)  \left(
D-3 \right) \left( D-2 \right) P_{2}(z)
\big],\nonumber\\
\nonumber\\
\beta_{3}&=& \frac{l^{2}\left[2\,{z}^{2}+\left(D-2\right)
\left(D-4\,z-1\right)\right]
}{4\, \left( D-3 \right)  \left( D-4 \right)  \left( z-D+2 \right) z}, \\ \nonumber\\
\Lambda&=&\frac{ \left( D-2 \right)  \left[ -{z}^{2}+ \left( D-2
\right) \left( D-1 -z \right)  \right]}{2\,
l^{2}\,\left(z-D+2\right)}.
\end{eqnarray*}
Note that this solution has been recently reported in \cite{Fan:2014ala}.As before the mass vanishes while the other relevant thermodynamical quantities are given by
\begin{subequations}
\begin{eqnarray}
S_{W}&=&{\frac {2 \,\pi\,\Omega_{D-2}\,\big(\Sigma_{2}\big)^{2}\,
r_{h}^{D-2}}{ \left( D-2 \right) \left( z-D+2 \right)\,l^{D-4} }}
 ,\\
T&=&{\frac { \left( D-2 \right) r_{h}^{z}}{2 \pi \,l^{z+1}}},\\
\mathcal{Q}&=&\Sigma_{2}\,\Omega_{D-2}\,\left(\frac{r_{h}}{l}\right)^{D-2},\\
\Phi&=&-{\frac {\Sigma_{2}\,r_{h}^{z}}{(z-D+2)\,l^{z-1}}},
\end{eqnarray}
\end{subequations}
in accordance with the results obtained in \cite{Fan:2014ala} via the Wald formalism. The extremal character of the charged solution is encoded in the relation $d \mathcal{M}=0=T d S_{W}+\Phi d \mathcal{Q}.$

\subsection{Third extremal solution with fixed $z=(D-1)$}
A third family of solutions is found when the dynamical exponent
takes the value $z=(D-1)$, and is given by
\begin{eqnarray}
ds^{2}_{D}&=&-\left(\frac{r}{l}\right)^{2\,(D-1)}f(r)\,dt^{2}+\frac{l^{2}}{r^{2}}\,\frac{dr^{2}}{f(r)}+\frac{r^{2}}{l^{2}}\,\sum_{i=1}^{D-2}
dx_{i}^{2}, \nonumber\\
f(r)&=&1-M\left(\frac{l}{r}\right)^{2(D-2)},\label{eq:metricsoln3}
\nonumber \\
F_{rt}&=& \sqrt{M}\,\Sigma_{3}, \label{eq:maxwellsoln3}
\end{eqnarray}
where
\begin{eqnarray*}
\big(\Sigma_{3}\big)^{2}&=&\big\{\left( D-2 \right)  \left[
\left(5\,{D}^{3}-20\,{D}^{2} +17\,D+4\right) l^{2}\right.\nonumber\\
&-&\left.2\, \left( 4\,D-7 \right) \left( D-3 \right)  \left( D-2
\right) ^{2}{\beta_2} \right] \big\}\nonumber\\
&\Big{/}&\big[ \kappa\,l^{4}\,\left( 5\,D-8 \right) \left( D-3
\right) \big],
\end{eqnarray*}
and the values of the coupling constants are given by
\begin{eqnarray*}
\beta_{1}&=&\frac {(2\,D-3) l^{2}-2\, \left( D-3 \right)  \left(
{D}^{2}-D-1 \right) \beta _{2}}{ 2 \left( D-1 \right)  \left( D-3
\right) \left( 5\,D-8 \right) }, \\ \nonumber\\
\beta_{3}&=& -\frac{l^{2}}{4\left(D-3\right)}, \\ \nonumber\\
\Lambda&=&-\frac{\left(D-2\right)\left(D-1\right)^2}{2\,l^{2}}.
\end{eqnarray*}
Note that in this case, the value of $\beta_2$ is not fixed, and this is the reason for which this solution differs from the previous one when $z=D-1$.
The Wald entropy and the Hawking temperature are described by
\begin{eqnarray}
S_{W}&=& {\frac {2 \pi
\,\Omega_{D-2}\,\big(\Sigma_{3}\big)^{2}\,r_{h}^{D-2}}{(D-2)\,l^{D-4}}}
,
\\
T&=&{\frac { \left( D-2 \right) r_{h}^{D-1}}{2\,\pi\,l^{D} }},
\end{eqnarray}
while the mass (\ref{eq:mastereq}) vanishes. The electric charge and potential are given by
\begin{eqnarray*}
\mathcal{Q}=\Sigma_{3}\,\Omega_{D-2}\,\left(\frac{r_{h}}{l}\right)^{D-2},\qquad
\Phi=-\frac{\Sigma_{3}\,r_{h}^{D-1}}{l^{D-2}},
\end{eqnarray*}
and once again, we verify that the first law of thermodynamics holds (\ref{ftlwa}) in its extremal form.

\subsection{Non extremal solution with $z=3(D-2)$}

Finally, a fourth family of solutions is found when the dynamical
exponent takes the value $z=3(D-2)$
\begin{eqnarray}
ds^{2}_{D}&=&-\left(\frac{r}{l}\right)^{6\,(D-2)}f(r)\,dt^{2}+\frac{l^{2}}{r^{2}}\,\frac{dr^{2}}{f(r)}+\frac{r^{2}}{l^{2}}\,\sum_{i=1}^{D-2}
dx_{i}^{2}, \nonumber\\
f(r)&=&1-M\left(\frac{l}{r}\right)^{2(D-2)},\label{eq:metricsoln4}
\nonumber \\
F_{rt}&=& \sqrt{M} \Sigma_{4} \,\left(\frac{r}{l}\right)^{2D-5},
\label{eq:maxwellsoln4}
\end{eqnarray}
where
\begin{eqnarray*}
\big(\Sigma_{4}\big)^{2}&=&\big\{ \left( D-2 \right)\left( D-3
\right) \big[ { l}^{2}
\left( 7\,D-13 \right)\left( 11\,D-27 \right)\nonumber\\
&-&8\, \left( 13 \,{D}^{2}-66\,D+86 \right) \left( D-2 \right)
^{2}{\beta_2} \big]  \left( D-4 \right)
 \big\}\nonumber\\
&\Big{/}&\big[\kappa\, Q_{3}(D)\,{l}^{4}\big],
\end{eqnarray*}
and the values of the coupling constants are fixed as
\begin{eqnarray*}
\beta_{1}&=& -\big[4\, \left( 15\,{D}^{2}-91\,D+142 \right)  \left(
D-2 \right) ^{3}{ \beta_2}\nonumber\\
&+&{l}^{2} \left( -17-38\,{D}^{2}+8\,{D}^{3}+53\,D \right)
\big]\nonumber\\
&\Big{/}& \big[4\, \left( D-2\right) ^{2} Q_{3}(D)
\big], \\ \nonumber\\
\beta_{3}&=& {\frac {{l}^{2} \left( 7\,D-13 \right) ^{2}-16\, \left(
D-1 \right)  \left( 2\,D-5 \right) \left( D-2\right) ^{2}{\beta_2}}{
 8\,\left( D-2\right) {Q_3(D)}}}
,
\\ \nonumber\\
\Lambda&=&-\big[96\, \left( 2\,D-5 \right)  \left( D-1 \right)
\left( D-3 \right)
 \left( D-4 \right)  \left( D-2 \right) ^{3}{\beta_2}\nonumber\\
 &+&{l}^{2}
 \left( D-2\right)  \left( 881\,{D}^{4}-8378\,{D}^{3}+30195\,{D}^{2}
\right.\nonumber\\
&-&\left.48626\,D+29384 \right) \big]\Big{/}
\big[4\,l^4\,Q_{3}(D)\big],
\end{eqnarray*}
where we have defined
\begin{equation*}
Q_{3}(D)=47\,{D}^{3}-369\,{D}^{2}+972\,D-848.
\end{equation*}
For this particular solution, the expression of the mass
(\ref{eq:mastereq}) is non-zero, and is given by
\begin{eqnarray}
\mathcal{M}&=&-\big\{\big[\big( 48\, \left( D-1 \right)  \left( D-3
\right) \left( D-4 \right)  \left( D-2 \right) ^{2} {\beta_2}
\nonumber\\
&+&{l}^{2} \left( 7\,D-13 \right)  \left( 13\,{D}^{2}-59\,D+76
\right)  \big)  \left( 2\,D-5 \right) \big]\nonumber\\
&\times& r_{h}^{4\left(D-2\right)} \Omega_{D-2}
\big\}\Big{/}\big[8\,\kappa\,Q_3(D)\,l^{4\,D-5}\big].
\label{massno0}
\end{eqnarray}
Note that this expression of the mass vanishes for the election
$$
\beta_2=-\frac{{l}^{2} \left( 7\,D-13 \right)  \left(
13\,{D}^{2}-59\,D+76 \right)}{48\, \left( D-1 \right)  \left( D-3
\right) \left( D-4 \right)  \left( D-2 \right) ^{2}},
$$
but this case reduces to the second family derived previously for a fixed value of the dynamical exponent
$z=3(D-2)$.

Calculating the Wald entropy, we obtain
\begin{eqnarray}
S_{W}&=& -\big\{\pi \,r_{h}^{D-2} \Omega_{D-2} \big[8\, \left(
25\,D-58 \right)  \left( D-3 \right)
\nonumber\\
&\times&\left( D-4 \right) \left( D-2 \right) ^{2}{\beta_2}+{l}^{2}
\left( 7\,D-13 \right)\nonumber\\
&\times& \left( 15\,{D}^{2}-49\,D+28 \right) \big] \big\}
\Big{/}\big[ \kappa \,l^{D}\,Q_3(D) \big] ,
\end{eqnarray}
while the Hawking temperature reads
\begin{eqnarray}
T&=&{\frac { \left( D-2 \right) r_{h}^{3\left(D-2\right)} } {2
\pi\,l^{3\,D-5}}}.
\end{eqnarray}
The electric thermodynamical quantities are given by
\begin{eqnarray}
\mathcal{Q}=\Sigma_{4}\,\Omega_{D-2}\left(\frac{r_{h}}{l}\right)^{D-2},\quad
\Phi=-{\frac
{\Sigma_{4}\,r_{h}^{3\left(D-2\right)}}{2\left(D-2\right)\,l^{3\,D-7}}},
\end{eqnarray}
and it is simple to check that the first law (\ref{ftlwa}) is satisfied.

\section{Lower-dimensional cases $D=3$ and $D=4$}

It is clear from the beginning that all these solutions are valid
for $D \geq 5$, due to the Gauss-Bonnet theorem in four dimensions
and the vanishing of the Gauss-Bonnet term in three dimensions. This
in turn implies that it is possible to switch off the contribution
of the coupling constant $\beta_{3}$ in four and three dimensions by
realizing the following shift
\begin{equation}
(\beta_{1},\beta_{2},\beta_{3})\rightarrow
(\beta_{1}-\beta_{3},\beta_{2}+4 \beta_{3},0).
\end{equation}

In three dimensions, among the four solutions derived previously,
only the last one can be projected in $D=3$ for a dynamical exponent
$z=3$ but in this case the solution is not longer charged since
$Q=0$. The resulting solution turns out to be the $z=3$ Lifshitz
black hole of new massive gravity \cite{AyonBeato:2009nh}. It is
interesting to note that in this case, the expression of the mass
(\ref{massno0}) becomes
$$
{\cal M}=-\frac{2\,\pi\, r_h^4}{\kappa\, l^4},
$$
and corresponds to the value of the mass found in \cite{Gonzalez:2011nz} with $\kappa=-8\pi G$.

Nevertheless, the third solution is regular in four dimensions and
becomes an extremal charged Lifshitz black hole with dynamical
exponent $z=3$ given by
\begin{eqnarray}
ds_4^2&=&-\left(\frac{r}{l}\right)^{6}
f(r)dt^2+\frac{l^{2}}{r^2\,f(r)}\,dr^{2}
+\frac{r^2}{l^{2}} \sum_{i=1}^{2}dx_{i}^2,\nonumber \\
f(r)&=&1-M\,\left(\frac{l}{r}\right)^{4},\nonumber \\
F_{rt}&=&2\,\sqrt{-\frac{3\,\beta_2\,M}{\kappa l^{4}}},
\end{eqnarray}
while the coupling constants take the form
\begin{eqnarray*}
\beta_{1}&=&\frac{l^{2}}{72}-\frac{11}{36}\,\beta_2 ,\qquad
\Lambda=-\frac{9}{l^{2}}.
\end{eqnarray*}

\section{Comments and conclusions}
The aim of this paper was to confirm in some concrete examples the validity of the recently proposed method to compute the quasilocal mass for arbitrary theory invariant under diffeomorphism and proposed in \cite{Kim:2013zha,Gim:2014nba}. This formalism is interesting for many reasons. Among other it does not require to linearize the field equations and yields to finite conserved charges independently of the asymptotic behavior of the metric solution. In our case, since we were interested on  charged Lifshitz black holes of the Einstein-Maxwell theory supplemented by the more general quadratic-curvature corrections, the quasilocal formalism fits perfectly with our intentions. Note that it is evident from the ADT formalism, the charged solutions derived here will have the same mass as their neutral part due to the fact that at the linearized level, the source terms corresponding to the Maxwell part can be put in the right hand side of the linearized equations \footnote{We thank B. O. Sarioglu to calling our attention in this point.}; see Refs. \cite{Devecioglu:2010sf} and \cite{Devecioglu:2011yi} for the computations of the masses within the ADT formalism in the neutral case.

Here, we have derived four classes of charged
Lifshitz black hole solutions where two of them do not require the dynamical exponent to be fixed. Three of these solutions are interpreted as extremal since the mass vanishes identically while the electric charge is non zero. In these cases, we have confirmed the validity of the first law of the thermodynamics. For the last family, the quasilocal mass and electric charge depend on the unique integration constant appearing in the solution, and as a matter of check we have also confirm that the first law holds in this case. Generically,
all these solutions are only valid for dimensions $D\geq 5$ because of the presence of the quadratic-curvature corrections. However, the solution with non-zero mass can be lowered to three dimensions, and in this case the dynamical exponent becomes $z=3$ while the electric charge vanishes, and the solution turns to be the Lifshitz $z=3$ black hole solution of new massive gravity \cite{AyonBeato:2009nh}. Interestingly enough, in this case, the expression of the mass becomes precisely the one of the Lifshitz $z=3$ black hole solution of new massive gravity \cite{Gonzalez:2011nz}.

We may notice that for our last solution with non zero mass and $z=3(D-2)$, the Smarr formula is given by
$$
{\cal M}=\frac{1}{4}\left(TS_W+\Phi{\cal Q}\right),
$$
and corresponds to the general Smarr formula derived in \cite{Dehghani:2013mba} and given by
$$
{\cal M}=\frac{D-2}{z+D-2}\left(TS_W+\Phi{\cal Q}\right).
$$
 It will be interesting to explore wether this Smarr formula is a consequence of some scaling symmetry of the reduced action.

\begin{acknowledgments}
We thank Eloy Ay\'on-Beato for useful discussions. MB is supported by BECA
DOCTORAL CONICYT 21120271. MH is partially supported by grant
1130423 from FONDECYT and from
CONICYT, Departamento de Relaciones Internacionales ``Programa
Regional MATHAMSUD 13 MATH-05''. This project was partially funded by Proyectos
CONICYT- Research Council UK - RCUK -DPI20140053.
\end{acknowledgments}

\section{Appendix}
In order to be self-contained and to clarify the most possible the draft, we report some of the formulas needed to derive our results.

\subsection{The fourth-order Cauchy differential equation arising from the combination $-\frac{l^{4}}{f}\,\left(
E_{t}^{t}-E_{r}^{r}\right)=0$}
\begin{eqnarray}
&&{r}^{3} \left( 2\, {\beta_3}+2\,{\beta_1}+{\beta_2} \right) \left[
r\,\frac{d^{4} f}{dr^{4}} +2\, \left( z+D+1
\right)\,\frac{d^{3} f}{dr^{3}} \right]\nonumber\\
&&+r^{2}\Big[\big( -2\,{z}^{2}+2\, \left( -D+8 \right) z+2\,
\left( D+5 \right)  \left( D-1 \right)  \big) \beta_{1} \nonumber\\
&&+ \big( -{z}^{2}+2\, \left( D+1 \right) z+{D}^{2}+2\,D-1
\big) \beta_{2} \nonumber\\
&&+\big( -2\,{z}^{2}+2\, \left( 5\,D-4 \right) z +2\,{D}^{2}+6\big)
\beta_{3}
\Big]\,\frac{d^{2}f}{dr^{2}}\nonumber\\
&&-(z+D-1)\,r \Big[2\, \left( z-1 \right)  \left( 2\,z+3\,D-5
\right) \beta_{1}\nonumber\\
&&+\left( 2\,{z}^{2}-z-D+1 \right) \beta_{2}+\left( 4\,{z}^{2}-6\,z
D \right. \nonumber\\
&&\left.+10\,z+2\,D-6 \right) \beta_{3} \Big]\frac{d f}{dr}- \Big[
\Big(  \left( 4\,{z}^{2}+8\,z-4\,z D+4
\right) \beta_{3}\nonumber\\
&&+ \left( 2\,D+2\,{z}^{2}-4 \right) \beta_{{2}}+ \left(
4\,{z}^{2}+4\,z D -8\,z+2\,{D}^{2}\right.\nonumber\\
&&-\left.6\,D+4 \right) \beta_{1} \Big) f -{l}^{2} \Big]
 \left( z-1 \right)  \left( D-2 \right)=0.\label{eq:cauchy}
\end{eqnarray}

\subsection{Expressions of  $\Xi_{1}$ and $\Xi_{2}$ appearing in Eq. (\ref{eq:ettii}) }
\begin{eqnarray*}
\Xi_{1}&=&-\big\{{l}^{2} \left[ {\chi}^{2}+ \left(
-3\,z+4-2\,D \right) \chi+2 \,{z}^{2}\right.\nonumber\\
&+&\left. \left( D-2 \right)  \left( D+2\,z-1 \right)  \right]
\left[ 2\, {\chi}^{3}  + \left( -4\,D+8 \right)
{\chi}^{2}\right.\nonumber\\
&+&\left. \big( -2\,{z}^{2}+ \left( 4\,D-8 \right) z + \left( D-1
\right) \left( D-4 \right)
\big) \chi\right.\nonumber\\
&-&\left.z \left( D-3 \right)  \left( D-4 \right)  \right] +\chi\, \left( D-3 \right)  \left( D-4 \right)\nonumber\\
&\times& \left( \chi+2-D -z \right)\big[ {\chi}^{3}+ \left(
-3\,D-2\,z+6 \right) {\chi}^{2}\nonumber\\
&+& \big( -{z}^{2}+ \left( 3\,D-6 \right) z +8+2\, D
\left( D-4 \right)  \big) \chi+2\,{z}^{3}\nonumber\\
&-&2\,z \left( z-2 \right) \left( D-2 \right)  \big] \beta_{2}
\big\}\nonumber\\
&\Big{/}&\big\{2\, \left( \chi+2-D-z \right) \chi\, \big[  \big(D
\left( D-3 \right) +4\, \left( D-2 \right) z \big) {\chi}^{2}\nonumber\\
&+& \big(  \left( -4\,D+8 \right) {z}^{2}+ \left(
19\,D-5\,{D}^{2}-16
 \right) z-  D \left( D-2 \right) \nonumber\\
 &\times& \left( D-3 \right)
 \big) \chi+ \left( 4\,D-8 \right) {z}^{3}+ \left( 16-2\,D-2\,{D}^{2
} \right) {z}^{2}\nonumber\\
&+&3\, \left( D-2 \right)  \left( {D}^{2}-3\,D+4
 \right) z-2\, \left( D-2 \right) \nonumber\\
 &\times& \left( {D}^{2}-3\,D+4 \right)
 \big] {l}^{4}
\big\},
\end{eqnarray*}

\begin{eqnarray*}
 \Xi_{2}&=&\big\{2\,{l}^{2} \big[ {\chi}^{4}+ \left( 4-D-2\,z
\right) {\chi}^{3}+\big( -{z}^{2}+ \left( 2\,z-3 \right) \nonumber\\
&\times& \left( D-2 \right)  \big) { \chi}^{2}+ \big( 2\,{z}^{3}+
\left( -3\,D+4 \right) {z}^{2} \nonumber\\
&+& \left( 4 \,D-6 \right) z-D \big) \chi+z \left( D-3 \right)
\left( D-4\right)  \left( -1+z \right)  \big] \nonumber\\
&\times& \big[ {\chi}^{2}+ \left( -3\, z+4-2\,D \right)
\chi+2\,{z}^{2}+ \left( 2\,D-4 \right) z\nonumber\\
&+& \left( D-1 \right)  \left( D-2 \right)  \big] +\chi\,
\left( D-3\right)  \left( D-4 \right)\nonumber\\
&\times&  \left( -D+2+\chi-z \right)  \big[ {\chi }^{4}+ \left(
-4\,z+6-2\,D \right) {\chi}^{3}\nonumber\\
&+& \big( 3\,{z}^{2}+ \left( 5\,D-14 \right) z+ \left( D-2 \right)
\left( D-6 \right)\big) {\chi}^{2}\nonumber\\
 &+& \big( 4\,{z}^{3}- D {z}^{2}-
 \left( D-2 \right)  \left( D-6 \right) z+2\, \left( D-2 \right) ^{2}
 \big) \chi\nonumber\\
 &-&4\,{z}^{4}+ \left( 4\,D-4 \right) {z}^{3}-4\,z \left( 3
\,z-2 \right)  \left( D-2 \right)  \big] \beta_{2} \big\}\nonumber\\
&\Big{/}&\big\{2\, \big[  \big(  \left( 4\,D-8 \right) z+ D \left(
D-3 \right)  \big) {\chi}^{2}+ \big(  \left( -4\,D
+8 \right) {z}^{2}\nonumber\\
&+& \left( 19\,D-5\,{D}^{2}-16 \right) z- D \left( D-2 \right)
\left( D-3 \right)  \big) \chi\nonumber\\
&+& \left( 4\,D-8 \right) {z}^{3}+ \left( 16-2\,D-2\,{D}^{2} \right)
{z}^ {2}+3\, \left( D-2 \right) \nonumber\\
&\times& \left( {D}^{2}-3\,D+4 \right) z-2\,
 \left( D-2 \right)  \left( {D}^{2}-3\,D+4 \right)  \big]\,\chi\, {l}^{4}
\big\}.
\end{eqnarray*}

\subsection{Expressions of  $\Psi_{1}$ and $\Psi_{2}$ appearing in Eq. (\ref{eq:mastereq}) }
\begin{eqnarray*}
\Psi_{1}&=&-(2\,{\beta_1}+{\beta_2}+2\,{\beta_3})\,\chi^{3}+\big[
\left( 2\,\beta_{{2}}+4\,\beta_{{3}}+4\,\beta_{{1}} \right) z
\nonumber\\
&+& \left( 2\,\beta_{2}+5\,\beta_{1}+3\,\beta_{3} \right) \left( D-2
\right)
\big]\,\chi^{2}\nonumber\\
&+&\big[\left( 2\,\beta_{1}+2\,\beta_{3}+\beta_{2} \right)
{z}^{2}-3\, \left( 3\,\beta_{3}+\beta_{1}+\beta_{2} \right)
\left(D-2 \right) z \nonumber\\
&-& \left( D-2 \right)  \big(  2 \left( 2\,D-3 \right) \beta_
{1}+\beta_{2} \left( D-2 \right) -2\,\beta_{3} \big)
\big]\,\chi \nonumber\\
&-&2 \left( 2\,\beta_{3}+\beta_{2}+2\,\beta_{1} \right) {z}^{3}-
\left( 2\,\beta_{1}-\beta_{{2}}-6\,\beta_{3} \right)
\left(  D-2 \right) {z}^{2} \nonumber\\
&-&2\, \left( 3\,\beta_{3}+\beta_{1}+\beta_{2} \right)  \left( D-2
\right) z+ \left( D-2 \right)
\nonumber\\
&\times& \big[  \left( D-2 \right)  \big( \beta_{1} \left( D-1
\right) +\beta_{{2}} \big) +2\,\beta_{3} \big], \\
\nonumber\\
\Psi_{2}&=& 2
\left(2\,\beta_{1}+\beta_{2}+2\,\beta_{3}\right)\,\chi^{3}-\big[ 4
\left( 2\,\beta_{3}+2\,\beta_{{1}}+\,\beta_{2} \right) z
\nonumber\\
&+&\left( 4\,\beta_{3}+8\,\beta_{1}+3\,\beta_{2} \right) \left(
D-2 \right) \big]\,\chi^{2}\nonumber\\
&+&\big[ -2\left( 2\,\beta_{3}+\,\beta_{2}+2\,\beta_{1} \right)
{z}^{2}+ 3\, \left( \beta_{2}+4\,\beta_{3} \right) \left( D-2
\right) z \nonumber\\
&+& \left( D-2 \right)  \big(  \left( D-2 \right) \beta_{{2}}+ 4
\left(
D-1 \right) \beta_{1}-4\,\beta_{3} \big) \big]\,\chi \nonumber\\
&+&4 \left( 2\,\beta_{3}+\beta_{2}+2\,\beta_{1} \right) {z}^{3}+2 \,
\left( 2\,\beta_{1}-\beta_{2}-6\,\beta_{3} \right)  \left( D-2
\right) {z}^{2} \nonumber\\
&+&4\, \left( 3\,\beta_{3}+\beta_{1}+\beta_{2} \right) \left( D-2
\right) z - \left[  2\left( D-2 \right) \beta_ {2}
\right.\nonumber\\
&+&\left.2\, \left( D-1 \right) \left( D-2 \right)
\beta_{1}+4\,\beta _{3}-l^{2} \right] \left( D-2 \right).
\end{eqnarray*}



\end{document}